\documentclass[aps,prd,epsf]{revtex4}

\begin{document} 

\title{
Boundary terms for supergravity and heterotic $M$-theory} 
\author{Ian G. Moss}
\email{ian.moss@ncl.ac.uk}
\affiliation{School of Mathematics and Statistics, University of  
Newcastle Upon Tyne, NE1 7RU, UK}

\date{\today}

%%%%%%%%%%%%%%%%%%%%%%%%%%%%%%%%%%%%%%%%%%%%%%

\begin{abstract}
This paper considers eleven dimensional supergravity on a manifold with
boundary and the theories related to heterotic $M$-theory, in which the matter
is confined to the boundary. New low energy actions and boundary conditions on
supergravity fields are derived. Previous problems with infinite constants
in the action are overcome. The new boundary conditions are shown to be
consistent with supersymmetry, and their role in the ten dimensional reduction
and gaugino condensation is briefly discussed.
\end{abstract}

\maketitle
%%%%%%%%%%%%%%%%%%%%%%%%%%%%%%%%%%%%%%%%%%%
\section{Introduction}

One of the interesting low energy limits of $M$-theory is thought to correspond
to eleven-dimensional supergravity with matter fields placed on two separate
ten-dimensional hypersurfaces. Horava and Witten have argued that this theory,
heterotic $M$-theory, describes the strongly coupled limit of the $E_8\times
E_8$ heterotic string in ten dimensions \cite{horava96,horava96-2}. The theory
can be compactified on a Calabi-Yau manifold to obtain a four dimensional
effective theory which is interesting from the point of view of particle
physics phenomenology \cite{witten96,banks96,lukas98,lukas98-2,ellis99}. 

The interaction terms in heterotic $M$-theory are constructed as an expansion
in $\kappa^{2/3}$, where $\kappa$ is the eleven dimensional gravitational
coupling constant. At leading order, the theory is simply eleven-dimensional
supergravity on the background $R^{10}\times S^1/Z_2$. Since the orbifold
part of the background $S^1/Z_2$ is identical to an interval $I$, the
background has a boundary consisting of two timelike ten-dimensional surfaces.
These provide the support for ten-dimensional gauge supermultiplets which
include Yang-mills and matter fields with Lagrangians which enter the model at
order $\kappa^{2/3}$. The gauge coupling constant and the gauge group $E_8$
are fixed by anomaly cancellation.

Previous attempts to construct an action have been hampered by the appearance
of the square of the Dirac delta function in the interactions at order
$\kappa^{4/3}$. This paper presents the details of an improved construction
which results in a consistent set of interaction terms up to order $\kappa^2$
\cite{moss03}. The main change is a modification of order $\kappa^{2/3}$ to
the boundary conditions on the gravitino and the supergravity three-form.
These changes to the boundary conditions play a similar role to the delta
function terms in the original Horava-Witten model, but remove the
singularities in the Horava-Witten model associated with having squares of the
delta function.

At leading order in $\kappa$, taking the gravitino to be a chiral field on the
boundary is consistent with the underlying eleven dimensional supergravity.
We shall see that corrections to the chirality condition of order
$\kappa^{2/3}$ are needed to ensure that the boundary condition remains
supersymmetric when the boundary matter is added into the picture. The
correction terms depend on the gauge field strength and a bilinear
combination of the gaugino.

The changes in the gravitino boundary condition are connected through the
supersymmetry to modifications to the supergravity three-form
boundary conditions depending on the gauge field strength and the gaugino.
These corrections to the boundary conditions are important when considering
the ten dimensional reduction of the eleven dimensional theory, where they
give rise to interaction terms between the gravity and Yang-Mills fields.
Furthermore, since gaugino condensation is a possible mechanism for
supersymmetry breaking in low energy heterotic $M$-theory
\cite{Horava:1996vs}, these corrections to the boundary conditions can be
particularly important when the supersymmetry is broken. 

The theory is presented here from the point of view of a manifold with
boundary, purely as a matter of technical convenience. The theory can also be
described on the covering space $R^{10}\times S^1$, where the boundary
condition on the gravitino can be viewed as a junction condition across the
hypersurface of fixed points of a $Z_2$ symmetry. The junction condition picks
up corrections from interaction terms between the gravitino and the matter
fields which live on the junction, and these corrections appear as the extra
matter terms in the boundary conditions.

The plan of this paper is as follows. The second section describes the
relationship between the covering space viewpoint and the manifold with
boundary picture, and provided the first indication that the gravitino boundary
condition has to be modified. The third section considers pure supergravity on
a manifold with boundary. The fourth section starts extends the discussion of
pure supergravity to include matter fields on the boundary, starting off with a
simplified description which neglects the four-fermi terms and then going into
technical detail to justify the full theory up to order $\kappa^2$. This
section ends with a brief discussion of the reduction of the theory to ten
dimensions. The following section shows that the action is supersymmetric up
to order $\kappa^2$. The conclusion
discusses some implications, particularly to supersymmetry breaking. 

In this paper the metric signature is $-+\dots+$. Eleven dimensional vector
indices are denoted by $I,J,\dots$. The coordinate indices on the boundary are
denoted by $A,B,\dots$ and  in the (outward unit) normal direction by $N$.
Eleven dimensional volume integrals are expressed in terms of 
$dv=|g|^{1/2}d^{11}x$,
where $|g|$ is the modulus of the determinant of the metric $g_{IJ}$. The
exterior derivative of an $n$-form $\alpha$ has components $({\rm
d}\alpha)_{I_1\dots I_{n+1}}=(n+1)^{-1}\partial_{[I_1}\alpha_{I_2\dots
I_{n+1}]}$ and the wedge product has components $(\alpha\wedge \beta)_{I_1\dots
I_m}={m\choose n}\,\alpha_{[I_1\dots I_n}\beta_{I_{n+1}\dots I_m]}$ where
$m\choose n$ is a binomial coefficient. The gamma matrices satisfy
$\{\Gamma_I,\Gamma_J\}=2
g_{IJ}$ and $\Gamma^{I\dots K}=\Gamma^{[I}\dots\Gamma^{K]}$. 
The spinors are Majorana, and
$\bar\psi=\psi^T\Gamma^0$.

\section{Junction conditions and boundary terms}

Junction conditions arise when we have field equations with sources which are
confined to hypersurfaces. In the situation where the hypersurface is fixed
under a reflection symmetry of the fields, it can easily be shown that the
junction conditions are equivalent to a set of boundary conditions. This gives
us two ways of describing the field theory: an `upstairs' distributional
description on the covering space or a `downstairs' description in terms of
fields and boundaries.

Consider the gravitational field with matter confined to a hypersurface
$\Sigma$ with surface stress-energy tensor $T_{AB}$. The Einstein equation
implies the
Israel junction conditions \cite{israel66,misner} on the extrinsic curvature,
\begin{equation}
\left[K_{AB}-g_{AB}K\right]_-=-\kappa^2 T_{AB},
\end{equation}
where $[f]_-$ denotes the change of a function $f$ across the surface and
$\kappa^2$ is the gravitational coupling constant.

Suppose that the hypersurface is fixed by the reflection symmetry $x\to {\cal
R}x$. In the neighbourhood of $\Sigma$, we can set up hypersurfaces $\Sigma_t$
which
are a distance $t$ from $\Sigma$ along the normal direction. The extrinsic
curvatures will transform by reflections according to
\begin{equation}
K_{AB}(x)=-K_{AB}({\cal R}x).
\end{equation}
In the limit $t\to 0$,
\begin{equation}
\left[K_{AB}-g_{AB}K\right]_+=0,
\end{equation}
where $[f]_+$ denotes the sum of the values of $f$ on either side of the
hypersurface. Consequently,
\begin{equation}
K_{AB}-g_{AB}K=\frac12\kappa^2 T_{AB}\label{pgbc}
\end{equation}
on the inside (ie the side from which the normal points) of the hypersurface.

The next example is a Rarita-Schwinger field with a distributional source
$J^A\delta_\Sigma$, where $\delta_\Sigma$ is a delta function. The source
appears in the Rarita-Schwinger equation
\begin{equation}
\Gamma^{AJK}D_J\psi_K=-J^A\delta_\Sigma.
\end{equation}
Integrating this along a direction normal to the hypersurface gives a junction
condition,
\begin{equation}
\Gamma^{AB}[\psi_B]_-=-J^A.
\end{equation}
We also have the $Z_2$ symmetry acting on the covering space,
\begin{equation}
\psi_A(x)=S\psi_A(Rx),
\end{equation}
where $S$ is a spinor transformation corresponding to the reflection symmetry.
We can choose this to be $S=\mp \Gamma_N$, where $\Gamma_N$ is the gamma matrix
associated with the normal direction. As the distance between $x$ and the
surface is reduced to zero, we obtain two alternative boundary
conditions
\begin{equation}
\Gamma^{AB}P_\pm\psi_B=\frac12J^A\label{rsbc}
\end{equation}
on the inside of $\Sigma$, where $P_\pm=\frac12(1\pm \Gamma_N)$. Either case
fixes exactly half of the fermion components on the boundary and gives a
complete set of boundary conditions for the Rarita-Schwinger equation. Note
that, for consistency, we also require $P_\mp J^A=0$.

This introduces one of the main points of this paper. The situation described
above applies to the low energy limit of the strongly coupled heterotic
string. Surface terms in the Lagrangian which include the gravitino lead to
surface sources and modifications to the gravitino boundary condition. The
chirality condition $\Gamma_{11}\psi_A=\psi_A$, which is often applied at the
boundary \cite{horava96}, should be modified to include extra terms. These
terms will be obtained later in the paper.

A similar argument applied to the normal component of the Rarita-Schwinger
field suggests that $P_\mp\psi_N=0$, because $\psi_N$ has the opposite
reflection parity to $\psi_A$. However, having fixed $\psi_A$, the boundary
conditions on $\psi_N$ can only be determined modulo a supersymmetry
transformation. The boundary conditions will therefore depend on the gauge
fixing condition which is used.

For the remainder of this section, we turn to the question of whether the
boundary conditions can be derived from an action principle. This will allow
us to use the `downstairs' description of a manifold with boundaries and avoid
the use of distributions. The action principle which generates the correct
gravitational junction conditions was found in \cite{hayward90}, and combined
with reflection symmetry to produce boundary conditions in reference
\cite{chamblin99}. Here we shall extend this to the
Rarita-Schwinger field.

Let ${\cal M}$ be a manifold with boundary, obtained by identifying the points
$x$ and ${\cal R}x$. On a manifold with boundary, the Einstein-Hilbert action
on ${\cal M}$ is supplemented by an extrinsic curvature term
\cite{york72,gibbons77}. We can also include some matter fields $\chi$ on the
boundary, with a surface Lagrangian ${\cal L}_s$. The natural candidate for the
total action is therefore
\begin{equation}
S_G=-{1\over\kappa^2}\int_{\cal M} R dv+
{2\over \kappa^2}\int_{\partial\cal M}K dv
+\int_{\partial\cal M}{\cal L}_s dv.
\end{equation}
Note that $\kappa$ denotes the Planck length on the covering space and this
results in a non-standard normalisation of the Einstein-Hilbert term on ${\cal
M}$.

The functional variation of the action arising from variations of the metric is
given by the results of appendix \ref{appa},
\begin{equation} 
\delta S_G=-{1\over\kappa^2}\int_{\cal M}G^{IJ}\delta g_{IJ}dv
-{1\over\kappa^2}\int_{\cal \partial M}
\left(K^{AB}-Kg^{AB}-\frac12\kappa^2T^{AB}\right)\delta g_{AB}dv.
\end{equation}
The action principle implies the Einstein equations $G_{IJ}=0$ and
the boundary condition (\ref{pgbc}).

The Rarita-Schwinger field can be introduced by including the action
\begin{equation}
S_{RS}= -{1\over\kappa^2}\int_{\cal M} 
\bar\psi_I\Gamma^{IJK} D_J(\omega)\psi_K dv,
\end{equation}
where $\omega$ is the tetrad connection (assumed torsion-free for the present).
The functional variation of the Rarita-Schwinger action includes a surface term
\begin{equation}
\delta S_{RS}=-{1\over\kappa^2}\int_{\cal \partial M}
\delta\bar\psi_A\Gamma^{AB} \Gamma_N\psi_B dv.
\end{equation}
If this was to vanish, the Rarita-Schwinger field equation would be
overconstrained. We can solve this problem by introducing an extra boundary
term (first used in \cite{luckock89}). The proposed form of the total action is
\begin{equation}
S={2\over\kappa^2}\int_{\cal M}
\left(-\frac12 R-\frac12\bar\psi_I\Gamma^{IJK} D_J(\omega)\psi_K\right)dv
+{2\over \kappa^2}\int_{\partial\cal
M}\left(K\mp \frac14\bar\psi_A\Gamma^{AB}\psi_B+
\frac{\kappa^2}2{\cal L}_s\right)dv,
\label{rsaction}
\end{equation}
where ${\cal L}_s\equiv{\cal L}(\chi,\psi_A)$ and either choice of sign is
allowed.

A useful technique, described in appendix B, allows us to replace tetrad
variations by metric variations. The surface terms in the variation of the
action are then
\begin{equation}
\delta S={2\over\kappa^2}\int_{\cal\partial M}
\left(\delta g_{AB}p^{AB}+\delta\bar\psi_A\theta^A\right)dv,
\label{vargrav}
\end{equation}
where
\begin{eqnarray}
\theta^A&=&\mp\Gamma^{AB}P_\pm\psi_B\pm\frac12J^A\label{taeq}\\
p^{AB}&=&-\frac12\left(K^{AB}-Kg^{AB}\right)+\frac14\kappa^2T^{AB}
\label{pabeq}
\end{eqnarray}
and the sources are,
\begin{eqnarray}
J^A&=&\pm\kappa^2{\partial{\cal L}_s\over\partial\psi_A}\\
T^{AB}&=&2{\partial{\cal L}_s\over\partial g_{AB}}+g^{AB}{\cal L}_s
\pm 2\bar\psi^{(A}\Gamma^{B)C}P_\pm\psi_C
\mp g^{AB}\bar\psi_C\Gamma^{CD}P_\pm\psi_D.
\end{eqnarray}
The action principle $\delta S=0$ gives the correct boundary conditions
(\ref{pgbc}) and (\ref{rsbc}) for both fields confirming that we have the
correct action. In general, the Rarita-Schwinger field gives a
contribution to the surface stress-energy tensor, although this contribution
vanishes if $P_\pm\psi_A=0$.

%%%%%%%%%%%%%%%%%%%%%%%%%%%%%%%%%%%%%%%%%%%%%%
\section{Boundary terms for supergravity}

In this section we shall construct the boundary terms for eleven-dimensional
supergravity. We shall be guided by the principle that the boundary conditions
should, as far as possible, be derivable from the extrema of the action. We
shall then show that the resulting action is supersymmetric. 

The fields are the metric $g$, gravitino $\psi_I$ and three-form $C$. The usual
supergravity action is \cite{cremmer78}
\begin{eqnarray}
S_{SG}=&&{2\over \kappa^2}\int_{\cal M}\left(-\frac12R(\Omega)
-\frac12\bar\psi_I\Gamma^{IJK}D_J(\Omega^*)\psi_K-\frac1{48}
G_{IJKL}G^{IJKL}\right.
\nonumber\\
&&\left.-\frac{\sqrt{2}}{192}\left(\bar\psi_I\Gamma^{IJKLMP}\psi_P
+12\bar\psi^J\Gamma^{KL}\psi^M\right)G^*_{JKLM}
-\frac{\sqrt{2}}{10!}
\epsilon^{I_1\dots I_{11}}(C\wedge G\wedge G)_{I_1\dots I_{11}}\right)dv,
\label{actionsg}
\end{eqnarray}
where $G$ is the abelian field strength and $\Omega$ is the tetrad connection.

The combination $G^*=(G+\hat G)/2$, where hats denote a standardised
subtraction of gravitino terms to make a supercovariant expression,
\begin{equation}
\hat G_{IJKL}=G_{IJKL}+\frac3{\sqrt{2}}\bar\psi_{[I}\Gamma_{JK}\psi_{L]}.
\end{equation}
Similarly, the combination $\Omega^*=(\Omega+\hat\Omega)/2$. If $\omega$ is the
Levi-Civita connection, then 
\begin{equation}
\hat\Omega_{IJK}=\omega_{IJK}
+\frac14\left(\bar\psi_I\Gamma_J\psi_K-\bar\psi_I\Gamma_K\psi_J
+\bar\psi_J\Gamma_I\psi_K\right).
\end{equation}
The connection $\Omega$ is given by
\begin{equation}
\Omega_{IJK}=\hat\Omega_{IJK}+\frac18\bar\psi^L\Gamma_{IJKLM}\psi^M.
\label{conf}
\end{equation}
In the usual 1.5 order formalism, the spin connection $\Omega$ is varied as
an independent field. Equation (\ref{conf}) results in the cancellation of the
terms which contain $\delta\Omega$. With the inclusion of the
boundary, we either have to modify equation (\ref{conf}), or to retain the
$\delta\Omega$ terms on the boundary. We shall keep the $\delta\Omega$ terms.

The boundary variation of the gravitational action must also include the
effects of torsion. The contorsion tensor ${\cal K}$ is defined by
\begin{equation}
\Omega_{IJK}=\omega_{IJK}+{\cal K}_{JIK}.
\end{equation}
The contorsion changes the Ricci scalar according to
\begin{equation}
R(\Omega)=R(\omega)+2D_J(\omega){\cal K}_I{}^{IJ}
-{\cal K}_I{}^{IK}{\cal K}_J{}^J{}_K-{\cal K}_{IJK}{\cal K}^{JKI}.
\end{equation}
The variation of the derivative leads to a boundary term. The simplest way to
cancel this term is to include the trace of the contorsion tensor ${\cal
K}_I{}^I{}_N$ in the boundary part of the action. When combined with the
action for the torsion-free theory (\ref{rsaction}), we are lead to the
boundary action 
\begin{equation}
S_0={2\over \kappa^2}\int_{\cal\partial M}\left(
K\mp\frac14\bar\psi_A\Gamma^{AB}\psi_B+
\frac12\bar\psi_A\Gamma^A\psi_N\right)dv.
\end{equation}
The two sign choices are equivalent to each other and only the plus sign will
be retained from now on. The full action $S=S_{SG}+S_0$.

It still remains to discuss the three-form field $C$. In the `upstairs'
description, the Bianchi identity leads to a junction condition
$[G_{ABCD}]_-=0$. However, the negative parity of the three-form field implies
that $[G_{ABCD}]_+=0$, consequently $G_{ABCD}=0$. This boundary condition is
not obtained from the variation of the action, in the same way that
the Bianchi identity is not one of the field equations obtained from
the action principle.

Now we are ready to check that the action $S$ is invariant under supersymmetry
transformations. To be more precise, we shall show that the supersymmetric
variation of the action vanishes after imposing the boundary conditions
$P_+\psi_A=0$ and $G_{ABCD}=0$.

The usual supersymmetry transformations for eleven-dimensional supergravity are
\begin{eqnarray}
\delta e^{\hat I}{}_J&=&\frac12\bar\eta\Gamma^{\hat I}\psi_{J}\label{varg}\\
\delta\psi_I&=&D_I(\hat\Omega)\eta+
{\sqrt{2}\over 288}\left(\Gamma_I{}^{JKLM}-8\delta_I{}^J\Gamma^{KLM}\right)
\eta\hat G_{JKLM}\label{varpsi}\\
\delta C_{IJK}&=&-{\sqrt{2}\over 8}\bar\eta\Gamma_{[IJ}\psi_{K]}\label{varc}
\end{eqnarray}
Some of the supersymmetry is broken by the boundary conditions $P_+\psi_A=0$
and $G_{ABCD}=0$, which are only preserved by supersymmetry transformations
which satisfy
\begin{equation}
P_+\eta=0
\end{equation}
on the boundary.

Consider first of all a general variation of the action, using the tetrad
formalism of appendix B,
\begin{eqnarray}
\delta S&=&{2\over\kappa^2}\int_{\cal M}dv\left(
\delta g_{IJ}E^{IJ}+\delta r_{IJ}Q^{IJ}+\delta\bar\psi_IL^I
+\delta C_{IJK}E^{IJK}\right)\nonumber\\
&&+{2\over\kappa^2}\int_{\cal \partial M}dv
\left(\delta g_{AB}\,p^{AB}+\delta\bar\psi_A\,\theta^A
+\delta C_{ABC}p^{ABC}
+\delta r_{AB}q^{AB}
\right).\label{varsg}
\end{eqnarray}
We will require explicit expressions for two of the boundary terms,
\begin{eqnarray}
\theta^A&=&-\Gamma^{AB}P_+\psi_B\\
p^{AB}&=&-\frac12\left(K^{AB}-Kg^{AB}\right)
+ 2\bar\psi^{(A}\Gamma^{B)C}P_+\psi_C
- g^{AB}\bar\psi_C\Gamma^{CD}P_+\psi_D
\end{eqnarray}
which follow from equations (\ref{taeq}) and (\ref{pabeq})

Now use the supersymmetry transformations (\ref{varg}-\ref{varc}). The volume
terms in the variation must cancell because of the invariance of the
supergravity action. Some additional boundary terms will arise from
integration by parts of the $\delta\psi_IL^I$ term, 
\begin{equation}
\delta S=
{2\over\kappa^2}\int_{\cal \partial M}dv
\left(\bar\eta L_N+\delta g_{AB}\,p^{AB}+\delta\bar\psi_A\,\theta^A
+\delta C_{ABC}p^{ABC}
+\delta r_{AB}q^{AB}
\right),
\end{equation}
After imposing the boundary condition $P_+\psi_A=0$ (which implies $\theta^A
=q^{AB}=\delta C_{ABC}=0$), we have
\begin{equation}
\delta S=
{2\over\kappa^2}\int_{\cal \partial M}dv
\left(\bar\eta L_N+\delta g_{AB}\,p^{AB}
\right).\label{varsgp}
\end{equation}
Let us examine the gravitino term more closely,
\begin{equation}
L_N=-\Gamma_N\Gamma^{AB}D_A(\hat\Omega)\psi_B
-\frac{\sqrt{2}}{96}\Gamma_N\Gamma^{ABCDE}\psi_A\hat G_{BCDE}
+\frac{\sqrt{2}}8\Gamma_N\Gamma^{AB}\psi^C\hat G_{ABCN}.\label{gfe}
\end{equation}
(The simplest way to obtain $L^I$ is to find the one-fermi terms in the
variation of the action and then use the fact that $L^I$ must be
supercovariant.) A slight rearrangement gives
\begin{equation}
\bar\eta L_N=\bar\eta D_A(\hat\Omega)\theta^A+
\frac12K_{AC}\bar\eta\Gamma^{AB}\Gamma^C\psi_B
-\frac{\sqrt{2}}{96}\bar\eta\Gamma^{ABCDE}\psi_A\hat G_{BCDE}
+\frac{\sqrt{2}}8\bar\eta\Gamma^{AB}\psi^C\hat G_{ABCN}.
\end{equation}
Imposing the boundary conditions $P_+\psi_A=0$ and $G_{ABCD}=0$, together with
the gamma-matrix identity (\ref{gmi1}), gives
\begin{equation}
\bar\eta L_N=
\frac12\left(K^{AB}-Kg^{AB}\right)\bar\eta\Gamma_A\psi_B.
\end{equation}
The two terms in equation (\ref{varsgp}) cancell and we conclude that the
action is supersymmetric.

We finish this section with some alternative ways to represent the boundary
action. In the first place, we can introduce the supercovariant form of the
extrinsic curvature,
\begin{equation}
\hat K^{AB}=
K^{AB}+\frac12\bar\psi^{(A}\Gamma^{B)}\psi_N+\frac14\bar\psi^A\psi^B,
\end{equation}
and the boundary action becomes
\begin{equation}
S_0={2\over \kappa^2}\int_{\cal\partial M}\left(
\hat K-\frac12\bar\psi_A\Gamma^A\Gamma^B\psi_B\right)dv.
\end{equation}
We can also improve the action by adding
\begin{equation}
S_c={2\over\kappa^2}\int_{\cal \partial M}dv \frac{\sqrt{2}}8 
C_{ABC}\bar\psi_D\Gamma^{DEABC}\psi_E.\label{sc}
\end{equation} 
The $\delta C_{ABC}p^{ABC}$ term in the variation of the total action then
takes a supercovariant form $\delta C_{ABC}\hat G^{ABCN}$. There is no need to
include $S_c$ if we impose the boundary conditions on the three-form when
varying the
action. In the following sections these restrictions will be used and the
boundary action $S_c$ omitted.

\section{Heterotic $M$-theory}

\subsection{Leading order\label{seca}}

Horava and Witten have argued that the low energy limit of the heterotic string
is given by eleven dimensional supergravity on the background $R^{10}\times
S^1/Z_2$, with $E_8$ gauge multiplets on the two ten dimensional fixed points
of the $Z_2$ symmetry, or the boundary branes as they are being described
here. The action is constructed by an expansion in powers of $\kappa^{2/3}$,
relying heavily on the restrictions of supersymmetry and the cancellation of
anomalies. A new gauge action and boundary conditions will be described in
this section at leading order, including terms with up to two fermion fields
and neglecting $R^2$ terms.

The gauge multiplets contain an $E_8$ gauge field $A^a_A$ and chiral fermions
$\chi^a$, which are in the adjoint representation. We begin with the
Super-Yang-Mills gauge action coupled to the Rarita-Schwinger field,
\begin{equation}
S'_1=-{2\epsilon\over\kappa^2}\int_{\cal \partial M}
\left(\frac14{F^a}_{AB}{F^a}^{AB}+\frac12\bar\chi^a\Gamma^AD_A(\Omega)\chi^a
+\frac14\bar\psi_A\Gamma^{BC}\Gamma^A{F^a}_{BC}\chi^a\right)dv.
\end{equation}
The constant $\epsilon$ sets the relative scale of the matter coupling. 

The usual supersymmetry transformations for gauge multiplet fields are
\begin{eqnarray}
\delta A^a{}_A&=&\frac12\bar\eta\Gamma_A\chi^a\\
\delta \chi^a&=&-\frac14\Gamma^{AB} F^a{}_{AB}\eta.
\end{eqnarray}
These are used allongside the rules for the supergravity fields
(\ref{varg}-\ref{varc}). 

In the original work of Horava and Witten, the supersymmetry transformation
rules where supplemented by extra terms containing distributions. Extra terms
also appeared in the gauge action. The combined effect lead to squares of
distributions at higher orders in $\kappa$. We argue here that modifications
have to be made to the boundary conditions and there is no need to modify the
supersymmetry transformations. Distributions never appear
explicitly, and problems with squares of distributions never arise.

In pure supergravity, the gravitino satisfied a chirality condition
$P_+\psi_A=0$ on the boundary. To leading order in fermion fields, the
supersymmetric variation of this chiral component is given by
\begin{equation}
\delta (P_+\psi_A)=P_+D_A\eta
+\frac{\sqrt{2}}{288}
\left(\Gamma_A{}^{BCDE}-8\delta_A{}^B\Gamma^{CDE}\right)
\eta G_{BCDE}.
\end{equation}
Assuming that $\eta$ has fixed chirality on the boundary implies that
$D_A(P_+\eta)=0$, and we have
\begin{equation}
P_+D_A\eta=-\frac12D_A\Gamma_N=\frac12K_{AB}\Gamma^B\eta.
\end{equation}
The extrinsic curvature is fixed by the stress energy tensor of the gauge
multiplet fields as in equation (\ref{pgbc}). To leading order in fermion
fields,
\begin{equation}
K_{AB}=F^a{}_A{}^CF^a{}_{BC}-\frac1{12}g_{AB}F^a{}^{CD}F^a{}_{CD}.
\end{equation}
The supersymmetric variation of the gravitino chirality condition to this order
in fermion fields is now
\begin{eqnarray}
\delta (P_+\psi_A)&=&
-\frac12\left(F^a{}_A{}^CF^a{}_{BC}-
\frac1{12}g_{AB}F^a{}^{CD}F^a{}_{CD}\right)\Gamma^B\eta\\
&&+\frac{\sqrt{2}}{288}
\left(\Gamma_A{}^{BCDE}-8\delta_A{}^B\Gamma^{CDE}\right)
\eta G_{BCDE}.\label{varchiral}
\end{eqnarray}
The boundary conditions $P_+\psi_A=0$ and $G_{ABCD}=0$ imply that $\delta
(P_+\psi_A)\ne 0$ for most choices of the non-abelian gauge field, breaking
the supersymmetry.

We would like to find a supersymmetric set of boundary conditions. In order to
be consistent with the boundary conditions for eleven dimensional supergravity
when $\epsilon=0$, we can deduce that any modifications to the boundary
conditions on the gravitino and three-form field should take the form
\begin{eqnarray}
P_+\psi_A&=&\epsilon f_A(\chi,A)\\
G_{ABCD}&=&\epsilon f_{ABCD}(\chi,A)
\end{eqnarray}
where $f_A$ is linear in $\chi$ and $f_{ABCD}$ contains terms up to quadratic
order in $\chi$. Gauge invariance and dimensional analysis restricts the
possible terms to `$F\chi$' combinations in $f_A$ and `$FF$' or `$\chi D
\chi$' terms in $f_{ABCD}$. By taking a linear combination of these terms it is
possible to show that 
\begin{eqnarray}
P_+\psi_A&=&{\epsilon\over
12}\left(\Gamma_A{}^{BC}-10\delta_A{}^B\Gamma^C\right)
F^a{}_{BC}\chi^a\\
G_{ABCD}&=&-3\sqrt{2}\epsilon F^a{}_{[AB} F^a{}_{CD]}
+\sqrt{2}\epsilon\bar\chi^a\Gamma_{[ABC}D_{D]}(\Omega)\chi^a\label{gbca}
\end{eqnarray}
is the unique supersymmetric combination.

We could impose a stronger condition on the three-form field by integrating
equation (\ref{gbca}),
\begin{equation}
C_{ABC}=-\frac{\sqrt{2}}{12}\epsilon\,\omega_{ABC}-
\frac{\sqrt{2}}{48}\epsilon\,\bar\chi^a\Gamma_{ABC}\chi^a
\end{equation}
where the Chern-Simons form
\begin{equation}
\omega={\rm tr}\left(A\wedge dA+\frac23A\wedge A\wedge A\right).
\end{equation}
We shall see in the next section that this boundary condition is still
supersymetric if we modify the transformation rule for $C_{ABC}$ to include an
abelian gauge transformation.

It is interesting to check consistency of the boundary condition on $C_{ABC}$
with the non-abelian gauge symmetry. Under a non-abelian gauge transformation
with $\delta A^a_A=-D_A\varepsilon^a$, the variation in the Chern-Simons form
becomes $\delta\omega=d(\varepsilon^aF^a)$. The non-abelian gauge
transformation of the Chern-Simons form can therefore be absorbed by an
abelian gauge transformation of the three-form (in analogy with
Yang-Mills Supergravity \cite{chapline83}). Let
\begin{equation}
\delta C_{ABC}=\frac{\sqrt{2}}{2}\epsilon\,\partial_{[A}a_{BC]},
\end{equation}
where $a_{AB}$ is an arbitrary two-form except that it must satisfy the
boundary condition
\begin{equation}
a_{AB}=\varepsilon^aF_{AB}^a.
\end{equation}
The action is unchanged, apart from a boundary term which comes from the
$C\wedge G\wedge G$ term in the supergravity action. This can be combined
with the quantum gauge anomaly to restore gauge invariance.

The mechanism described above is a simple variation of the generalised
Green-Schwarz \cite{green84} mechanism found by Horava and Witten
\cite{horava96-2}, and it fixes the the expansion parameter,
\begin{equation}
\epsilon={1\over 4\pi}\left({\kappa\over4\pi}\right)^{2/3}.
\end{equation}
Furthermore, extending the argument to gravitational and mixed anomalies
suggests that the $F\wedge F$ term in (\ref{gbca}) should be replaced by 
$F\wedge F-\frac12R\wedge R$,  but we shall drop the $R^2$ terms from further
discussion for the present.

The boundary conditions were chosen for consistency with the supersymmetry
transformations. We shall now find that the gravitino boundary
condition can be derived completely independently from the extrema of the
boundary action given at the beginning of this section. The total action is
given by $S=S_{SG}+S'_1$, where $S_{SG}$ is
the supergravity action and $S'_1$ is the boundary term.

We vary the tetrad as described in appendix A. The boundary terms in the
variation of the total action $S$ are then
\begin{equation}
\delta S={2\over\kappa^2}\int_{\cal \partial M}dv
\left(\delta g_{AB}\,p^{AB}+\delta\bar\psi_A\,\theta^A
+\delta C_{ABC}p^{ABC}
+\delta\bar\chi\,\Xi+\delta A_AY^A
\right),\label{varsp}
\end{equation}
Variation of the gravitino field in $S'_1$ is elementary, and combines
with the supergravity results from the previous section to give
\begin{equation}
\theta^A=-\Gamma^{AB}P_+
\psi_B-\frac{\epsilon}{4}\Gamma^{BC}\Gamma^A F^a{}_{BC}\chi^a.
\end{equation}
The boundary condition $\theta^A=0$ can be manipulated using the gamma-matrix
identities in appendix C into the previous form
\begin{equation}
P_+\psi_A={\epsilon\over 12}
\left(\Gamma_A{}^{BC}-10\delta_A{}^B\Gamma^C\right)F_{BC}\chi.
\end{equation}
The remaining terms in the variation of the action give the graviton boundary
condition, as in equation (\ref{pgbc}), and the field equations for $\chi^a$
and $A^a_A$. 

The action is supersymmetric subject to the boundary conditions on the
gravitino and the antisymmetric gauge field. For example, the variation of the
`$F\psi\chi$' term in the surface action produces expressions of the form
`$F^2\psi\eta$' and `$GF\chi\eta$'. These terms cancel with `$G\psi\eta$'
terms from the variation of the supergravity action when we apply the boundary
conditions on $G$ and $\psi$. (In the original Horava-Witten model, an extra
term in the boundary action was needed to cancel the `$GF\chi\eta$' variation.)
A more detailed discussion of the supersymmetric invariance of the action will
be given in a later section.

%%%%%%%%%%%%%%%%%%%%%%%%%%%%%%%%%%%%%%%%%%%
\subsection{Boundary conditions}

In this section we shall construct a supersymmetric set of boundary conditions
on the gravitino and three-form field which include all of the fermion terms
(but still ignoring the $R^2$ terms).
The most important new feature is the appearance of bilinear gaugino terms, not
only in the gravitino and three-form boundary conditions, but also in the
boundary conditions on the supersymmetry parameter.

The supersymmetry transformations of the fields will be almost as before,
\begin{eqnarray}
\delta A^a{}_A&=&\frac12\bar\eta\Gamma_A\chi^a\\
P_-\delta \chi^a&=&-\frac14\Gamma^{AB}\hat F^a{}_{AB}\eta\label{vchi}
\end{eqnarray}
The field strength is now in supercovariantised form
\begin{equation}
\hat F^a{}_{IJ}=F^a{}_{IJ}-\bar\psi_{[I}\Gamma_{J]}\chi.
\end{equation}
We now have to allow for the possibility that $P_+\delta\chi\ne 0$. This can be
accomodated as described in appendix B. 

The boundary condition on the tangential components of the three-form
which we saw in the previous section can be imposed as a constraint
$c_{ABC}=0$, where 
\begin{equation}
c_{ABC}=C_{ABC}+\frac{\sqrt{2}}{12}\epsilon\,\omega_{ABC}+
\frac{\sqrt{2}}{48}\epsilon\,\bar\chi^a\Gamma_{ABC}\chi^a.\label{cbc}
\end{equation}
By dimension counting, this boundary condition cannot contain
four-fermi terms unless we go to higher orders in the expansion parameter
$\epsilon$.

We need to pay special attention to the supersymmetric variation of the
Chern-Simons form. This decomposes into a gauge invariant part and a total
derivative, 
\begin{equation}
\delta\omega=2\delta A^a\wedge F^a-d(A^a\wedge\delta A^a)
\end{equation}
We will require that $c_{ABC}$ should be invariant under supersymmetry
transformations modulo abelian gauge transformations. We could absorb the
abelian transformation by modifying the supersymmetry transformation,
\begin{equation}
\delta C_{IJK}=-{\sqrt{2}\over 8}\bar\eta\Gamma_{[IJ}\psi_{K]}
+\frac{\sqrt{2}}4\epsilon\partial_{[I}f_{JK]},
\end{equation}
where the two-form $f$ can be chosen arbitrarily except that it must
satisfy the boundary condition
\begin{equation}
f_{AB}=2A^a{}_{[A}\delta A^a{}_{B]}.\label{agt}
\end{equation}
This modification would have no effect on the variation of gauge invariant
terms in the action.

The one-fermi terms in the gravitino boundary condition which were evaluated in
the previous section suggest that a suitable anzatz for the gravitino boundary
condition would be
\begin{equation}
P_+\psi_A={\epsilon\over 12}\left(\Gamma_A{}^{BC}-10\delta_A{}^B\Gamma^C\right)
\hat F^a{}_{BC}\chi^a-\epsilon\Gamma P_-\psi_A.\label{anz}
\end{equation}
The term involving $\Gamma$, where $\Gamma$ is a `$\chi\chi$' bilinear, is
allowed on dimensional grounds. Note that if we apply $P_-$ to the anzatz we
find that $P_-\Gamma=\Gamma P_+$.

The gravitino anzatz could also be put into a more suggestive form
\begin{equation}
\tilde P_+\psi_A={\epsilon\over
12}\left(\Gamma_A{}^{BC}-10\delta_A{}^B\Gamma^C\right)
\hat F^a{}_{BC}\chi^a.
\end{equation}
where a new projection operator is defined by
\begin{equation}
\tilde P_+=P_++\epsilon\Gamma P_-.
\end{equation}
This also suggests that we should consider a modification to the supersymmetry
parameter, and impose
\begin{equation}
\tilde P_+\eta=0.\label{ebc}
\end{equation} 
on the boundary.

Now we are ready to examine the variation of the three terms in $c_{ABC}$ under
supersymmetry transformations. (We shall omit the abelian gauge transformation
(\ref{agt}).) Firstly, using the transformation (\ref{varc}), equations
(\ref{anz}), (\ref{ebc}) and gamma-matrix identities
\begin{equation}
\delta C_{ABC}=
-\frac{\sqrt{2}}{96}\epsilon\bar\eta\Gamma_{ABC}{}^{DE}\chi\hat F_{DE}
+\frac{\sqrt{2}}{16}\epsilon\bar\eta\Gamma_{[AB}{}^D\chi\hat F_{C]D}
-\frac{3\sqrt{2}}{16}\epsilon\bar\eta\Gamma_{[A}\chi\hat F_{BC]}
+\frac{\sqrt{2}}8\epsilon\bar\eta\{\Gamma,\Gamma_{[AB}\}P_-\psi_{C]}
\end{equation}
Secondly, the variation of the Chern-Simons terms gives
\begin{equation}
\delta\omega_{ABC}=3\bar\eta\Gamma_{[A}\chi\hat F_{BC]}
+3\bar\eta\Gamma_{[A}\chi\,\bar\chi\Gamma_B\psi_{C]}
\end{equation}
up to an abelian gauge transformation. Thirdly, the gaugino term produces
\begin{equation}
\delta(\bar\chi\Gamma_{ABC}\chi)=
\frac12\bar\eta\Gamma_{ABC}{}^{DE}\chi\hat F_{DE}
-3\bar\eta\Gamma_{[AB}{}^D\chi\hat F_{C]D}
+\frac32\bar\eta\Gamma_D\psi_{[A}\bar\chi\Gamma^D{}_{BC]}\chi.
\end{equation}

We may combine the three expressions together and use a Fierz rearrangement
(\ref{fbc}) to get
\begin{equation}
\delta c_{ABC}=
-\frac{\sqrt{2}}{768}\epsilon\bar\chi\Gamma_{DEF}\chi\,
\bar\eta\{\Gamma^{DEF},\Gamma_{[AB}\}P_-\psi_{C]}
+\frac{\sqrt{2}}8\epsilon\bar\eta\{\Gamma,\Gamma_{[AB}\}P_-\psi_{C]}.
\end{equation}
From this we can read off
\begin{equation}
\Gamma=\frac1{96}\bar\chi\Gamma_{ABC}\chi\,\Gamma^{ABC}.\label{geq}
\end{equation}
With this choice, the boundary condition on the three-form field is
supersymmetric modulo an abelian gauge transformation, with no approximations. 
Note that $\tilde P_+\chi=P_+\chi=0$ due to a convenient Fierz identity.

Taking the exterior derivative of $c_{ABC}=0$ puts the boundary condition in
manifestly gauge invariant (and supercovariant) form,
\begin{equation}
\hat G_{ABCD}=-3\sqrt{2}\epsilon\hat F^a{}_{[AB}\hat F^a{}_{CD]}
+\sqrt{2}\epsilon\bar\chi^a\Gamma_{[ABC}D_{D]}(\hat\Omega)\chi^a
+\frac{\sqrt{2}}{4}\epsilon\bar\chi^a\Gamma_{[ABC}\Gamma^{EF}\psi_{D]}\hat
F^a{}_{EF}.\label{gbc}
\end{equation}
Again, this is an exact result (neglecting $R\wedge R$ terms) and improves on
the approximate result given in the previous section. The derivation of
(\ref{gbc}) requires use of the gravitino boundary condition,  providing a
check on the expression for $\Gamma$ in (\ref{geq}).

The supersymmetric variation of the gravitino boundary condition can also be
examined along the lines of equation (\ref{varchiral}). It is easy to confirm
that the boundary condition is supersymmetric up to at least the one-fermi
terms.
%%%%%%%%%%%%%%%%%%%%%%%%%%%%%%%%%%%%%%%%%%
\subsection{The Lagrangian}

The boundary terms in the action can be constructed by imposing the basic
requirement that the extrema of the action should generate the boundary
conditions of the previous sections. There is one exception, which is the
boundary condition $c_{ABC}=0$ on the three-form field, which has to be
imposed separately. The boundary conditions on the gravitino field, in
particular, restrict the possible four-fermi terms in the Lagrangian. In the
next section, it will be shown that the action is supersymmetric up order
$\epsilon^3$. 

Consider the following boundary action to replace the boundary action of
section \ref{seca},
\begin{equation}
S_1=-{2\epsilon\over\kappa^2}\int_{\cal \partial M}dv
\left(\frac14{F^a}_{AB}
{F^a}^{AB}+\frac12\bar\chi^a\Gamma^AD_A(\hat\Omega)\chi^a
+\frac14\bar\psi_A\Gamma^{BC}\Gamma^A{F^a}^*_{BC}\chi^a
+\frac1{192}\bar\chi\Gamma_{ABC}\chi\bar\psi_D\Gamma^{ABCDE}\psi_E\right),
\label{action1}
\end{equation}
where $F^*=(F+\hat F)/2$. The coefficient of the four-fermi term has been
chosen with some fore-knowledge. This coefficient depends on whether we use
$\hat\Omega$ or $\Omega^*$ in the gaugino derivative.

The total action
\begin{equation}
S=S_{SG}+S_0+S_1.
\end{equation}
At the extrema of the action, it is possible to read off the field equations
and the boundary conditions. We shall examine these in more detail in the next
section.

As a check, consider the variation of the action due to a variation of the
gravitino field. The surface term
\begin{eqnarray}
\delta_\psi S_1&=&-{2\epsilon\over \kappa^2}\int_{\cal M}dv
\left(\frac14\delta\bar\psi_A\Gamma^{BC}\Gamma^A F^*{}^a{}_{BC}\chi^a
-\frac18\bar\psi_A\Gamma^{BC}\Gamma^A\chi^a\delta\bar\psi_B\Gamma_C\chi^a
\right.\nonumber\\
&&+\left.\frac1{16}\delta\bar\psi_A\Gamma_B\psi_C
\bar\chi^a\Gamma^{ABC}\chi^a
+\frac1{96}\delta\bar\psi_A\Gamma^{ABCDE}\psi_B
\bar\chi^a\Gamma_{CDE}\chi^a
\right).
\end{eqnarray}  
By a Fierz rearrangement (\ref{fga}),
\begin{eqnarray}
\delta_\psi S_1&=&-{2\epsilon\over \kappa^2}\int_{\cal M}dv
\left(\frac14\delta\bar\psi_A\Gamma^{BC}\Gamma^A \hat F^a{}_{BC}\chi^a
+\frac1{16}\delta\bar\psi_A\Gamma_B\psi_C
\bar\chi^a\Gamma^{ABC}\chi^a\right.\nonumber\\
&&+\left.\frac1{192}\delta\bar\psi_A[\Gamma^{AB},\Gamma^{CDE}]\psi_B
\bar\chi^a\Gamma_{CDE}\chi^a
+\frac1{96}\delta\bar\psi_A\Gamma^{ABCDE}\psi_B
\bar\chi^a\Gamma_{CDE}\chi^a
\right)
\end{eqnarray}
Using the gamma-matrix identities (\ref{gmi1}-\ref{gmi4}),
\begin{equation}
\delta_\psi S_1=-{2\epsilon\over \kappa^2}\int_{\cal M}dv
\left(\frac14\delta\bar\psi_A\Gamma^{BC}\Gamma^A \hat F^a{}_{BC}\chi^a
+\frac1{96}\delta\bar\psi_A\Gamma^{AB}\Gamma^{CDE}\psi_B
\bar\chi^a\Gamma_{CDE}\chi^a.
\right)
\end{equation}
This can be combined with the supergravity result to get the $\theta^A$ term in
equation (\ref{vargrav}),
\begin{equation}
\theta^A=-\Gamma^{AB}P_+
\psi_B-\frac{\epsilon}{4}\Gamma^{BC}\Gamma^A\hat F^a{}_{BC}\chi^a
-\frac{\epsilon}{96}\Gamma^{AB}\Gamma^{CDE}\psi_B\bar\chi^a\Gamma_{CDE}\chi^a.
\end{equation}
The boundary condition $\theta^A=0$, obtained by variation of the action, can
be rewritten using gamma-matrix identities as equation (\ref{anz}). We recover
the boundary condition which was derived from supersymmetry in the previous
section.

%%%%%%%%%%%%%%%%%%%%%%%%%%%%%%%%%%%%%%%%%%%%%%%%%%%%

\subsection{Ten dimensional reduction}

This is a good place to consider the relationship between the new
11-dimensional theory and $N=1$ Yang-Mills supergravity in ten dimensions. We
shall see how some important features of the ten-dimensional theory are
explained by their eleven dimensional precursor. In order to keep the
description manageable, we shall not go into the full details of the reduction
from eleven to ten dimensions, but focus rather on the important features.

The general procedure for reducing the eleven dimensional theory is identical
to the reduction of the Horava-Witten model \cite{Lukas:1997rb,Lukas:1998ew}. 
This reduction depends on two small parameters. In the first place we have the
parameter $\epsilon$ in the matter action. The reduction also assumes that the
length scales characterising the variation of the fields in ten dimensions are
much larger than the brane separation $L$. Since the matter fields appear in
the boundary conditions, the bulk fields will in general depend on the
eleventh dimension. The first step in the reduction is to solve the field
equations for the bulk to order $\epsilon$ with the matter fields held
constant. This forms the beginning of a perturbative solution with small
$\epsilon$ and small ten-dimensional derivatives. The $O(\epsilon^0)$ terms
give the ten-dimensional supergravity action and additional terms give rise to
the matter couplings.

Consider one of the boundary surfaces ${\cal \partial M}^1$ placed at
$x^{11}=0$ and another ${\cal \partial M}^2$ at $x^{11}=L$. In this section,
we shall use $n$ to denote the outgoing normal to the surface at  $x^{11}=L$.
To leading order, we can take the metric to be
\begin{equation}
g=e^{-2\phi}g'_{AB}dx^Adx^B+e^{4\phi/3}dx^{11}dx^{11}
\end{equation}
with 10-dimensional metric $g'_{AB}$ and dilaton $\phi$. However, for this
brief account it is sufficient to express results in terms of 11-dimensional
metric and gamma matrix components.

A feature of new boundary Lagrangian (\ref{action1}) is that contains no
`$\chi\chi G$' term. This term plays an important role in Horova-Witten theory
where it combines with other terms to form a perfect square involving the
3-form field strength $H_{ABC}$ in the 10-dimensional reduction
\cite{Horava:1996vs}. In the new theory, we turn instead to the boundary
conditions (\ref{cbc}),
\begin{eqnarray}
C_{ABC}&=&+\frac{\sqrt{2}}{12}\epsilon\,\tilde\omega^1{}_{ABC}
\hbox{   on   } {\cal \partial M}^1\\
C_{ABC}&=&-\frac{\sqrt{2}}{12}\epsilon\,\tilde\omega^2{}_{ABC}
\hbox{   on   }{\cal \partial M}^2
\end{eqnarray}
where 
\begin{equation}
\tilde\omega_{ABC}=\omega_{ABC}+
\frac{1}{4}\,\bar\chi^a\Gamma_{ABC}\chi^a
\end{equation}
The change of sign at the boundary surface ${\cal \partial M}^1$ is due to the
normal vector $n$ being an ingoing normal there rather than an outgoing
normal.

The bulk solution is determined by the boundary conditons, the Bianchi identity
(with no source terms) and the divergence of $G$. In our approximation, the
general solution is
\begin{eqnarray}
C_{ABC}&=&-\frac{\sqrt{2}}{12}\epsilon y\,
\tilde\omega^{(2)}{}_{ABC}
+\frac{\sqrt{2}}{12}\epsilon (1-y)\,
\tilde\omega^{(1)}{}_{ABC}\\
C_{11AB}&=&\frac16 B_{AB}
\end{eqnarray}
where $y=x^{11}/L$ and  $B_{AB}$ is a constant of integration which becomes our
ten dimensional 2-form field with field strength related to $H_{ABC}$.
Terms in the ten-dimensional theory depending on $B_{AB}$ come from the field
strength components $G_{11ABC}$,
\begin{equation}
G_{11ABC}=3\partial_{[A}B_{BC]}-{\sqrt{2}\over 2}{\epsilon\over L}
\left(\tilde\omega^1_{ABC}+\tilde\omega^2_{ABC}\right),
\end{equation}
The $H_{ABC}$ terms appear as a perfect square in the Lagrangian
due to the term $G_{11ABC}G^{11ABC}$ in the 11-dimensional action. This
reproduces the same low energy behavior as the Horava-Witten theory, but in
our case no modification of the field strength (singular or non-singular) is
involved.

We can also see how the $H_{ABC}$ fields and the gaugino enter some of the
supersymmetry transformation rules of the ten-dimensional theory. Consider the
dilatino $\psi_{11}$ with 11-dimensional transformation
\begin{equation}
\delta\psi_{11}=D_{11}(\hat\Omega)\eta-
\frac{\sqrt{2}}{36}\Gamma^{ABC}\hat G_{11ABC}\eta+
\frac{\sqrt{2}}{24}\Gamma^{ABCD}\hat G_{ABCD}\eta
\end{equation}
and keep only $B_{AB}$ and $\chi^a$ non-zero. We have to find a way of
evaluating $D_{11}(\hat\Omega)\eta$. The boundary conditions (\ref{ebc}) for
$\eta$ are
\begin{eqnarray}
P_+\eta&=&+\frac{\epsilon}{96}\Gamma^{ABC}\bar\chi^{1a}\Gamma_{ABC}\chi^{1a}
P_-\eta\hbox{   on   }{\cal \partial M}^1\\
P_+\eta&=&-\frac{\epsilon}{96}\Gamma^{ABC}\bar\chi^{2a}\Gamma_{ABC}\chi^{2a}
P_-\eta\hbox{   on   }{\cal \partial M}^2.
\end{eqnarray}
At leading order in our expansion the supersymmetry parameter must be the same
as the 10-dimensional supersymmetry parameter $\eta'$. At order $\epsilon$ the
supersymmetry parameter has to depend on $x^{11}$ in order to satisfy the
boundary conditions. We can choose
\begin{equation}
\eta=\eta'-\frac{\epsilon}{96}
\Gamma^{ABC}\left(y\bar\chi^{2a}\Gamma_{ABC}\chi^{2a}
-(1-y)\bar\chi^{1a}\Gamma_{ABC}\chi^{1a}\right)\eta'
\end{equation}
where $P_+\eta'=0$. Consequently, the $D_{11}(\hat\Omega)\eta$ term depends on
the gaugino field. The dilatino transformation becomes
\begin{equation}
\delta\psi_{11}=-\frac{\sqrt{2}}{36}\Gamma^{ABC}H_{ABC}\,\eta'
-\frac1{256}{\epsilon\over L}\Gamma^{ABC}
\bar\chi^{a'}\Gamma_{ABC}\chi^{a'}\,\eta'
\end{equation}
where $\chi^{a'}=(\chi^{1a},\chi^{2a})$. 

This reduction of the 11-dimensional theory explains why the combination of
$H_{ABC}$ and $\chi^a$ terms which appear in the 10-dimensional action
as a perfect square does not also appear in the dilatino supersymmetry
transformation. The reason for the difference can be traced back to the strange
extra bilinear terms (\ref{geq}) which we found in the supersymmetry parameter
(and gravitino) boundary condition at the three-fermi level.

Other bulk fields have similar behaviour to the ones mentioned above. For
example, if we solve the 11-dimensional gravitino equation with the boundary
conditions (\ref{gbca}) to the same order of approximation which we used
above, and also neglect the three-fermi terms, then
\begin{eqnarray}
\psi_A&=&\psi'_A-
{\epsilon\over 12}
\left(\Gamma_A{}^{BC}-10\delta_A{}^B\Gamma^C\right)
\left(yF^{2a}{}_{BC}\chi^{2a}+(1-y)F^{1a}{}_{BC}\chi^{1a}\right)
\label{phianz}\\
\psi_{11}&=&\psi_{11}'+\epsilon\psi^{(1)}\label{dilanz}
\end{eqnarray}
where $\psi_A'$ and $\psi_{11}'$ are 10-dimensional chiral fields, 
$P_+\psi'_A=P_-\psi_{11}'=0$. The background spinor field $\psi^{(1)}$ is a
solution to
\begin{equation}
\Gamma^{AB}D_B\psi^{(1)}=-\frac14\Gamma^{BC}\Gamma^A
F^{a'}{}_{BC}\chi^{a'}.
\end{equation}
The existence of solutions to this equation, at least locally, is made possible
by an integrability
condition which is satisfied as a consequence of the conservation of the
Yang-Mills supercurrent. 

The 10-dimensional gravitino is a linear combination of $\psi'_A$ and
$\psi'_N$. Inserting the fields (\ref{phianz}) and (\ref{dilanz}) into the
11-dimensional action leads to the usual supergravity action at leading order
\cite{green}. Interaction terms involving  `$\chi F\psi_A$' arise from the
order $\epsilon$ terms in (\ref{phianz}). Beyond order $\epsilon$, the
background spinor field $\psi^{(1)}$ starts to appear in the action and may
cause difficulties, but this remains to be investigated.

%%%%%%%%%%%%%%%%%%%%%%%%%%%%%%%%%%%%%%%%%%%%
\section{Supersymmetry}

We turn now to the supersymmetric variation of the action. We shall make use of
the supersymmetric invariance of the 11-dimensional supergravity action
without boundaries and follow a similar route to the one used earlier in
section III. We shall see below how the most of the two-fermi terms in the
supersymmetric variation of the action cancel. Only one term remains, which is
of order $\kappa^2$. An outline of the treatment of the four-fermi terms is
given in appendix D.

To begin, consider the general variation of the action rather than a
supersymmetric one. Using the tetrad formalism of appendix B, we are able to
write a general variation of the action in the form
\begin{eqnarray}
\delta S&=&{2\over\kappa^2}\int_{\cal M}dv\left(
\delta g_{IJ}E^{IJ}+\delta r_{IJ}Q^{IJ}+\delta\bar\psi_IL^I
+\delta C_{IJK}E^{IJK}\right)\nonumber\\
&&+{2\over\kappa^2}\int_{\cal \partial M}dv
\left(\delta g_{AB}\,p^{AB}+\delta\bar\psi_A\,\theta^A
+\delta C_{ABC}p^{ABC}
+\delta\bar\chi\,\Xi+\delta A_AY^A
+\delta r_{AB}q^{AB}
\right),\label{vars}
\end{eqnarray}
The coefficients appearing here are the field equations and boundary
conditions, given explicitly by
\begin{eqnarray}
p^{AB}&=&-\frac12\left(\hat K^{AB}-\hat Kg^{AB}\right)
+\frac14\kappa^2 \hat T^{AB}\label{gravfe}\\
p^{ABC}&=&\hat G^{ABCN}+
\frac{\sqrt{2}}{4}\bar\psi_D\Gamma^{DEABC}P_+\psi_E
-\frac{\sqrt{2}}{3\times 7!}\epsilon^{ABCD_1\dots D_7}(C\wedge G)_{D_1\dots
D_7}\label{cnfe}\\
\Xi&=&\epsilon\left(-\Gamma^AD_A(\hat\Omega)\chi-
\frac14\Gamma^A\Gamma^{BC}\hat F_{BC}\psi_A\right)\label{chife}\\
Y^A&=&\epsilon\left(-D_B(\omega)\hat F^{AB}+
\frac12D_B(\omega)(\bar\psi_C\Gamma^{BAC}\chi)
+\frac18\bar\psi_D\Gamma^{ABCDE}\psi_E F_{BC}\right)
\end{eqnarray}
The variation of the metric is the most complicated. This has been simplified
by assuming that $\hat T^{AB}$ is in supercovariant form,
\begin{eqnarray}
\frac14\kappa^2 \hat T^{AB}&=&
\epsilon\left(
\frac12\hat F^a{}^{AC}\hat F^a{}^B{}_C-
\frac18g^{AB}\hat F^a{}^{CD}\hat F^a{}_{CD}
+\frac14\bar\chi^a\Gamma^A D^B(\hat\Omega)\chi^a
-\frac14g^{AB}\bar\chi^a\Gamma^C D_C(\hat\Omega)\chi^a\right.\nonumber\\
&&+\left.\frac1{16}\bar\chi^a\Gamma^A\Gamma^{CD}\psi^B\hat F^a{}_{CD}
-\frac1{16}\bar\chi^a\Gamma^C\Gamma^{DE}\psi_C\hat F^a{}_{DE}g^{AB}
\right)
\end{eqnarray}
Furthermore, the condition $\delta c_{IJK}=0$ has been used to simplify
equation (\ref{cnfe}). This is equivalent to the addition of equation
(\ref{sc}).

The supersymmetric variation can be found by substituting the
supersymmetry transformations (\ref{varg}-\ref{varc}) into
the general expression for the variation of the action. 
We know that the volume terms in the variation cancell because
of the invariance of the supergravity action without boundaries. As in section
III, additional boundary terms arise from integration by parts of the
$\delta\psi_IL^I$ term. Terms which vanish when
$\theta^A=0$ can be dropped, leaving
\begin{equation}
\delta S={2\over\kappa^2}\int_{\cal \partial M}dv
\left(\bar\eta L_N+\delta g_{AB}\,p^{AB}
+\delta C_{ABC}p^{ABC}
+\delta\bar\chi\,\Xi+\delta A_AY^A+\delta r_{AB}q^{AB}
\right),\label{varsa}
\end{equation}

The global supersymmetry of the Super-Yang-Mills action is related to the
identity
\begin{equation}
\delta A_A(-D_B(\hat\Omega)\hat F^{AB})+
\delta\bar\chi\Gamma^AD_A(\hat\Omega)\chi
=\frac12\bar\eta D_A(\hat\Omega)J^A-
\frac14\bar\eta\Gamma^{ABC}\chi\,D_A(\hat\Omega)(\bar\psi_B\Gamma_C\chi),
\end{equation}
where $J^A$ is the supercovariantised supercurrent of the gauge multiplet.
Inserting this into $\delta S$ and dropping the four-fermi terms gives 
\begin{equation}
\delta S={2\over\kappa^2}\int_{\cal \partial M}dv
\left(\bar\eta L_N+\delta g_{AB}\,p^{AB}
+\delta C_{ABC}p^{ABC}
+\frac12\bar\eta D_A(\hat\Omega)J^A
-\epsilon\delta\bar\chi\Gamma^A\Gamma^{BC}\hat F_{BC}\psi_A\right).
\label{varsb}
\end{equation}

The expression of $L_N$ was given in an earlier section (\ref{gfe}). We can
rearrange $\bar\eta L_N$ into the suggestive form
\begin{eqnarray}
\bar\eta L_N&=&\bar\eta D_A(\hat\Omega)\theta^A+
\frac12\left(K^{AB}-Kg^{AB}\right)\bar\eta\Gamma_A\psi_B
-\frac12\bar\eta D_A(\hat\Omega)J^A
\nonumber\\
&&-\frac{\sqrt{2}}{96}\bar\eta\Gamma^{ABCDE}\psi_A\hat G_{BCDE}
+\frac{\sqrt{2}}8\bar\eta\Gamma^{AB}\psi^C\hat G_{ABCN}\nonumber
\end{eqnarray}
There are also additional four-fermi contributions from the
modification to the projection operator $\tilde P_+$.

The $\delta\chi$ term can be replaced by using the identity
\begin{equation}
\delta\bar\chi\Gamma^A\Gamma^{BC}\hat F_{BC}\psi_A=
\frac14\bar\eta\Gamma^{ABCDE}\psi_A\hat F_{BC}\hat F_{DE}
-2\bar\eta\Gamma_A\psi_B
\left(\hat F^{CA}\hat F_C{}^B-\frac14g^{AB}\hat F^{CD}\hat F_{CD}\right)
\end{equation}
which follows from the gamma-matrix identities (\ref{gmi1}-\ref{gmi4}). 

Putting these together, the two-fermi terms in the supersymmetric variation of
the action are
\begin{eqnarray}
\delta S&=&{2\over\kappa^2}\int_{\cal \partial M}dv
\left(\delta g_{AB}\,\left(p^{AB}+\frac12(K^{AB}-Kg^{AB})-
\left(\hat F^{CA}\hat F_C{}^B-
\frac14g^{AB}\hat F^{CD}\hat F_{CD}\right)\right)
\right.\nonumber\\
&&\left.+\delta C_{ABC}\left(p^{ABC}-\hat
G^{ABCN}\right)-\frac{\sqrt{2}}{96}\bar\eta\Gamma^{ABCDE}\psi_A
\left(\hat G_{BCDE}+3\sqrt{2}\hat F_{AB}\hat F_{CD}\right)\right).
\end{eqnarray}
Examination of the field equations (\ref{gravfe}-\ref{chife}) and the boundary
condition (\ref{gbc}) shows immediately that most of the two-fermi terms in
$\delta S$ cancel. The only term remaining comes from $p^{ABC}$ and it is
\begin{equation}
\delta S={2\over\kappa^2}\int_{\cal \partial M}dv\,2
\frac{\sqrt{2}}{10!}\epsilon^{A_1\dots A_{10}}
(\delta C\wedge C\wedge G)_{A_1\dots A_{10}}.
\end{equation}
Since $\delta C_{ABC}$, $C_{ABC}$ and $G_{ABCD}$ are all of order $\epsilon$,
the variation $\delta S$ is of order $\epsilon^3$, or equivalently $\kappa^2$.

\section{conclusion}

The strongly coupled limit of the heterotic string is believed to be  related
to eleven-dimensional supergravity on a manifold with boundary. The main
results of this paper have been the construction of a consistent set of
boundary conditions and a corresponding
supersymmetric action, including the effects of gauge fields on the boundary. A
major missing ingredient so far has been the $R^2$ terms, but otherwise the
theory is a possible candidate for heterotic $M$ theory.

The main differences between the new theory and Horava-Witten  theory 
are the following:
\begin{enumerate}
\item Terms involving the square of the Dirac delta function do not occur in
the new theory.
\item The gravitino boundary condition $\Gamma_{11}\psi_A=\psi_A$ has been
replaced by the boundary condition
\begin{equation}
\tilde P_+\psi_A={\epsilon\over
12}\left(\Gamma_A{}^{BC}-10\delta_A{}^B\Gamma^C\right)
\hat F^a{}_{BC}\chi^a,\label{gbcagain}
\end{equation}
where $\tilde P_+=P_++\epsilon\Gamma P_-$ is a modified projection operator
which depends on the gaugino expectation value $\Gamma$, and the constant
$\epsilon=O(\kappa^{2/3})$ by a modification of the usual anomaly cancellation
argument. The new boundary condition is supersymmetric, consistent with
junction conditions across the brane and consistent with boundary variations
of the action. 

\item The new boundary Lagrangian contains no `$\chi\chi G$' term. This term
plays an important role for the Horova-Witten theory where it combines with
other terms to form a perfect square in the 10-dimensional reduction
\cite{Horava:1996vs,Lukas:1997rb}. In the new theory, 
the boundary conditions 
\begin{equation}
\hat G_{ABCD}=-3\sqrt{2}\epsilon\hat F^a{}_{[AB}\hat F^a{}_{CD]}
+\sqrt{2}\epsilon\bar\chi^a\Gamma_{[ABC}D_{D]}(\hat\Omega)\chi^a
+\frac{\sqrt{2}}{4}\epsilon\bar\chi^a\Gamma_{[ABC}\Gamma^{EF}\psi_{D]}\hat
F^a{}_{EF}
\end{equation}
lead to a similar effect.
\end{enumerate}

The differences between the low energy theory obtained here
and the original formulation of Horava and Witten leads to some
modifications to the effects of gaugino condensation. The gaugino condensate
appears in the supersymmetry transformations of the dilatino, as it does in
Horava-Witten theory. In the new formulation,
the gaugino condensate also appears in the gravitino boundary conditions,
through the bilinear expression $\Gamma$ defined in equation (\ref{geq}).
To examine the effects of the gaugino terms in the graviton boundary condition,
suppose that the two boundaries are at $x^{11}=0$ and $x^{11}=L$, with a
gaugino condensate on $x^{11}=L$. If the expectation values of linear gaugino
terms vanishes, then
the gravitino boundary condition ({\ref{gbcagain}) becomes
\begin{eqnarray}
P_+\psi_A&=&0,\qquad x^{11}=0\\
(P_-+\epsilon\Gamma P_+)\psi_A&=&0, \qquad x^{11}=L\label{tgbc}
\end{eqnarray}
The fermion
boundary condition on $x^{11}=L$ can be put into a more familiar
form when the condensate has the property that the matrix $\Gamma=\sigma I$,
where $I^2=-1$. In this case this boundary condition becomes
\begin{equation}
(1+e^{\theta I}\Gamma_N e^{-\theta I})\psi_A=0,
\end{equation}
where $\tan\theta=\epsilon\sigma$.
These boundary conditions are themselves sufficient to break the supersymmetry,
and they have been introduced before, usually in contexts which were
unconnected to gaugino condensation
\cite{Antoniadis:1997ic,fabinger00,flachi01,gersdorff02,brax02}. We are able to
conclude that, in heterotic $M$-theory,  $\theta$ depends on the magnitude of
the gaugino condensate. 
(A similar condition is also induced by the $\kappa$-symmetry of a $D$ brane,
where instead of depending on the condensate, $\theta$ depends on the Born
Infeld-field \cite{bergshoeff97}.)  

The results have been obtained in the very lowest energy regime where higher
derivative terms can be neglected. This is not a consistant approximation if
we would like to reduce the theory to lower dimensions
\cite{witten96,banks96,lukas98,lukas98-2,ellis99}. The $R^2$ terms in
the action and the Lorentz Chern-Simons form can simply be added in a way which
reduces to the correct 10-dimensional theory.  One direction in which the
current discusion needs to be improved is to include these terms in a
supersymmetric fashion on the boundary in 11 dimensions, but this is a
complicated task. 

\acknowledgments

I have benefitted from helpful discussions about gaugino condensation with
James Norman. 

\appendix

\section{Metric variations\label{appa}}

Standard formulae for the variation of the curvature and the extrinsic
curvature which have been used in the text have been collected together in
this appendix. Further details can be found in
\cite{misner,hayward90,chamblin99}. Let $n^I$ be the outward normal to the
boundary. The intrinsic
metric and extrinsic curvatures of the boundary are
defined by
\begin{equation}
h_{IJ}=g_{IJ}-n_In_J
\end{equation}
and
\begin{equation}
K_{IJ}=h_{IK}h_{JL}n^{K;L},
\end{equation}
where $;I$ denotes the components of the covariant derivative using the
Levi-Civita connection. Note that $h_{AB}=g_{AB}$.

The variation of the Ricci scalar $R$ is given by \cite{misner}
\begin{equation}
\delta R=g^{IJ}g^{KL}\left(\delta g_{IJ;KL}-\delta g_{IK;JL}\right)
-R^{IJ}\delta g_{IJ}
\end{equation}
Variation of the volume integral of $R$ gives a boundary term which can be
decomposed into,
\begin{equation}
n^Ig^{JK}\left(\delta g_{JK;I}-\delta g_{IJ;K}\right)
=-h^{JK}(n^I \delta h_{IJ} )_{;K}+K^{IJ}\delta h_{IJ}
+h^{JK}n^I\delta h_{JK;I}-2n^IK\delta n_I.
\label{dricci}
\end{equation}
The only restriction which has been imposed is that $\delta(n^In_I)=0$. The
variation of the trace of the extrinsic curvature can be arranged into
\begin{equation}
\delta K=-h^{JK}(n^I\delta h_{IJ})_{;K}+\frac12n^Ih^{JK}\delta h_{JK;I}
-n^IK\delta n_I.
\end{equation}
These two variations can be combined to give
\begin{equation}
-\int_{\cal M}dv\,\delta R+2\int_{\partial\cal M}dv\,\delta K=
\int_{\cal M}dv\,R^{IJ}\delta g_{IJ}-
\int_{\partial\cal M}dv\,K^{IJ}\delta h_{IJ},
\end{equation}
asuming that the boundary consists of smooth components.

\section{Tetrad variations}

This appendix explains how the variation of the action with respect to the
tetrad can be simplified by making use of Lorentz invariance. It particular, it
will be shown that it is possible to vary the action
in a restricted way which is analagous to fixing the connection in the 1.5
order formalism.

We first decompose the tetrad variation into a metric variation and a Lorentz
transformation. Since the metric 
\begin{equation}
g_{IJ}=e^{\hat I}{}_I e_{\hat IJ},
\end{equation}
we have
\begin{equation}
\delta g_{IJ}=\delta e^{\hat I}{}_I e_{\hat IJ}+
e^{\hat I}{}_I \delta e_{\hat IJ}
\end{equation}
We now define the antisymmetric combination to represent Lorentz
transformations
\begin{equation}
\delta r_{IJ}=\delta e^{\hat I}{}_I e_{\hat IJ}-
e^{\hat I}{}_I \delta e_{\hat IJ}.
\end{equation}
An arbitrary tetrad variation can then be decomposed according to the equation
\begin{equation}
\delta e^{\hat I}{}_I=\frac12(\delta g_{IJ}+\delta r_{IJ})e^{\hat I J}.
\end{equation}
The variation of the tetrad connection $\omega_{I\hat J\hat K}=e_{\hat
J}\cdot D_I e_{\hat K}$, is given by
\begin{equation}
\delta\omega_{I\hat J\hat K}=\delta g_{I[\hat J;\hat K]}
-\delta r_{\hat J\hat K;I}.
\end{equation}
We can make use of these relations to replace the tetrad variations in the
variation of the action by $\delta g_{IJ}$ and $\delta r_{IJ}$.

Consider a Lagrangian ${\cal L}\equiv{\cal L}(e_{\hat I},\psi_I)$ for
fields on a manifold ${\cal M}$ with no boundary. The
general variation of the action takes the following form
\begin{equation}
\delta S=\int_{\cal M}\left(\delta g_{IJ}E^{IJ}+\delta r_{IJ}Q^{IJ}
+\delta\bar\psi_I \theta^I\right)dv
\end{equation}
If we restrict the variation to a local Lorentz transformation of the tetrad
$e_{\hat I}$ and the spinors, then the action must be unchanged, hence
\begin{equation}
0=\int_{\cal M}\left(\delta r_{IJ}Q^{IJ}
-\frac18\delta r_{IJ}\bar\psi_K\Gamma^{IJ}\theta^K
\right)dv.
\end{equation}
We therefore deduce that
\begin{equation}
Q^{IJ}=\frac18\bar\psi_K\Gamma^{IJ}\theta^K.
\end{equation}
Now consider the introduction of a boundary $\partial {\cal M}$ and a fermion
field $\chi$ on the boundary. In an adaptive coordinate system $e_{\hat N
I}=\delta_{N I}$, and the terms allowed in the boundary part of the variation
are
\begin{equation}
\delta S=\int_{\partial\cal M}dv\left(\delta g_{AB}p^{AB}+\delta r_{AB}q^{AB}
+(\delta g_{AN}+\delta r_{AN})\,q^{AN}
+\delta\bar\psi_I \theta^I+\delta\bar\chi\,\Xi\right).
\end{equation}
Local Lorentz invariance implies
\begin{equation}
q^{IJ}=\frac18\bar\psi_K\Gamma^{IJ}\theta^K
+\frac18\bar\chi\Gamma^{IJ}\Xi.\label{qij}
\end{equation}
These results allow a considerable reduction in the amount of effort required
to find the variation of the action. The variation can be done initially with
$\delta r_{AB}=0$, and then the $\delta r_{AB}$ terms can be
recovered from the fermion variations. Similarly, if $\theta^A=0$ and $\chi$ is
a chiral fermion, then $q^{AN}=0$ and the variation of the action can be
obtained by setting $\delta g_{AN}=\delta r_{AN}=0$.

The tetrad variations also affect the variation of the gaugino. The chirality
condition $P_+\chi=0$ implies that
\begin{equation}
P_+\delta\chi=-(\delta P_+)\chi=
-\frac12\Gamma^A(\delta g_{AN}+\delta r_{AN})\label{pmindchi}
\end{equation}
However, when $q^{AN}=0$ these terms must cancel with other terms in the
variation of the action. Similarly, $\delta g_{AN}$ can be ignored in the
boundary variation of the contorsion terms in the Ricci tensor.

\section{Spinor identities}

The gamma-matrix conventions used in this paper are
\begin{equation}
\{\Gamma_I,\Gamma_J\}=2 g_{IJ},\qquad
\Gamma^{I_1\dots I_n}=\Gamma^{[I_1}\dots\Gamma^{I_n]}.
\end{equation}
The covariant derivative of a spinor $\zeta$ is
\begin{equation}
D_I(\omega)\zeta=\partial_I\zeta+\frac14\omega_{IJK}\Gamma^{JK}\zeta.
\end{equation}
All of the spinors are Majorana and we have
\begin{equation}
\bar\epsilon\,\Gamma^{I_1\dots I_n}\,\eta=
=\sigma_n\,\bar\eta\,\Gamma^{I_1\dots I_n}\,\epsilon.
\end{equation}
where $\sigma_r=(-1)^{r(r+1)/2}=+,-,-,+$ for $r=0,1,2,3$ mod 4.

Products of gamma-matrices can be expanded by the following identities, valid
in $d$ dimensions,
\begin{eqnarray}
\Gamma^{I_1\dots I_n}\Gamma_{J_1\dots J_m}&=&
\sum_{r=0}^{min(n,m)}\pmatrix{n\cr r\cr}\pmatrix{m\cr r\cr}
(-1)^{nr}\sigma_r r!\delta^{[I_1}_{[J_1}\dots\delta^{I_r}_{J_r}
\Gamma^{I_{r+1}\dots I_n]}{}_{J_{r+1}\dots J_m]},\\
\Gamma^{I_1\dots I_n}\Gamma_{J_1\dots J_m}\Gamma_{I_1\dots I_n}&=&
\sum_{r=0}^{min(n,m)}\pmatrix{d-m\cr n-r\cr}\pmatrix{m\cr r\cr}
(-1)^{r+(m-1)n}\sigma_n \Gamma_{J_1\dots J_m}.
\end{eqnarray}
In ten dimensions, products of gamma-matrices which have
been used most often in this paper are
\begin{eqnarray}
\Gamma^{AB}\Gamma_{C}&=&\Gamma^{AB}{}_{C}
-2\delta_C{}^{[A}\Gamma^{B]}\label{gmi1},\\
\Gamma_{C}\Gamma^{AB}&=&\Gamma^{AB}{}_{C}
+2\delta_C{}^{[A}\Gamma^{B]},\\
\Gamma^{ABC}\Gamma_{DE}&=&\Gamma^{ABC}{}_{DE}
+6\delta_{[D}{}^{[A}\Gamma_{E]}{}^{BC]}-
6\delta_{[D}{}^{[A}\delta_{E]}{}^B\Gamma^{C]},\\
\Gamma_{DE}\Gamma^{ABC}&=&\Gamma^{ABC}{}_{DE}
-6\delta_{[D}{}^{[A}\Gamma_{E]}{}^{BC]}-
6\delta_{[D}{}^{[A}\delta_{E]}{}^B\Gamma^{C]}.\label{gmi4}
\end{eqnarray}
Some other examples are given in table \ref{table1}.

\begin{table}[h]
\caption{\label{table1} Coefficients in the contraction 
$\Gamma^{I_1\dots I_n}\Gamma_{J_1\dots J_m}\Gamma_{I_1\dots I_n}
=c_{nm}\Gamma_{J_1\dots J_m}$ in 10 dimensions.}
\begin{tabular}{|l|r|r|r|r|r|r|} \hline 
  $c_{nm}$         & $m=0$ &$m=1$ & $m=2$&$m=3$& $m=4$&$m=5$\\ \hline 
$n=1$&10&-8&6&-4&2&0\\
$n=2$&-90&-54&-26&-6&6&5\\
$n=3$&-720&288&-48&-48&48&0\\
$n=4$&5040&1008&-326&306&48&5\\
$n=5$&30240&0&-3660&0&1440&0\\
\hline
\end{tabular}
\end{table}

Ten dimensional Fierz rearrangements can be obtained from the formula
\begin{equation}
\zeta \bar\chi=-\frac1{16}\bar\chi\Gamma_A\zeta\,\Gamma^A
+\frac1{96}\bar\chi\Gamma_{ABC}\zeta\,\Gamma^{ABC}
-\frac1{3840}\bar\chi\Gamma_{ABCDE}\zeta\,\Gamma^{ABCDE}.\label{fierz}
\end{equation}
valid for chiral spinors $\Gamma_{11}\chi=\chi$ and $\Gamma_{11}\zeta=\zeta$
(and also for pairs of antichiral spinors), where
$\Gamma_{11}=\Gamma_1\dots\Gamma_{10}$ . The following examples make use of
the Fierz identity (\ref{fierz}) and the product formulae, 
\begin{eqnarray}
\delta\bar\psi_A\Gamma^{BC}\Gamma^A\chi\,\bar\psi_B\Gamma_C\chi
-\delta\bar\psi_B\Gamma_C\chi\,\bar\psi_A\Gamma^{BC}\Gamma_A\chi
&=&4\bar\chi\Gamma_{CDE}\chi\,
\delta\bar\psi_A[\Gamma^{AB},\Gamma^{CDE}]\psi_B\label{fga}
\\
\bar\chi\Gamma_{D[AB}\chi\,\bar\eta\Gamma^D\psi_{C]}
+\frac16\bar\chi\Gamma_{DEF}\chi\,\bar\eta\Gamma_{[AB}{}^{DEF}\psi_{C]}
&=&-16\bar\eta\Gamma_{[A}\chi\,\bar\chi\Gamma_B\psi_{C]}
\label{fbc}\\
\bar\zeta\Gamma^A\Gamma_{BC}\psi_A\bar\chi\Gamma_B\psi_C
-\bar\zeta\Gamma_B\psi_C\bar\chi\Gamma^A\Gamma^{BC}\psi_A&=&
-2\bar\zeta\Gamma^B\chi\,\bar\psi_B\Gamma^A\psi_A
\end{eqnarray}
where $\chi$ and $\zeta$ are chiral spinors, but $\psi_A$ is not assumed
chiral.

%%%%%%%%%%%%%%%%%%%%%%%%%%%%%%%%%%%%%%%%%%
\section{Supersymmetry of the action}

The version of equation (\ref{varsb}) which describes the supersymmetric
variation of the action and includes the four fermi terms is
\begin{eqnarray}
\delta S&=&{2\over\kappa^2}\int_{\cal \partial M}dv
\left(\bar\eta L_N+\delta g_{AB}\,p^{AB}+\delta r_{AB}q^{AB}
+\delta C_{ABC}p^{ABC}
+\frac12\bar\eta D_A(\hat\Omega)J^A
-\epsilon\delta\bar\chi\Gamma^A\Gamma^{BC}\hat F_{BC}\psi_A\right.\nonumber\\
&&\left. 
+\frac14\epsilon\bar\eta\Gamma_B\chi\,D_A(\hat\omega)
(\bar\psi_C\Gamma^{ABC}\chi)
-\frac14\epsilon\bar\eta\Gamma^{ABC}\chi\,D_A(\hat\Omega)
(\bar\psi_B\Gamma_C\chi)
+\frac12\epsilon\bar\eta\Gamma_A\chi\,D_B(\hat{\cal K})\hat F^{AB}
\right).
\end{eqnarray}
where $D(\hat{\cal K})=D(\hat\Omega)-D(\omega)$. 

The full expression for $\bar\eta L_N$ is
\begin{eqnarray}
\bar\eta L_N&=&\bar\eta D_A(\hat\Omega)\theta^A+
\frac12\left(K^{AB}-Kg^{AB}\right)\bar\eta\Gamma_A\psi_B
-\frac12\bar\eta D_A(\hat\Omega)J^A
\nonumber\\
&&-\frac{\sqrt{2}}{96}\bar\eta\Gamma^{ABCDE}\psi_A\hat G_{BCDE}
+\frac{\sqrt{2}}8\bar\eta\Gamma^{AB}\psi^C\hat G_{ABCN}\nonumber\\
&&+\epsilon\bar\eta \Gamma^{AB}(D_A(\hat\Omega)\Gamma)\psi_B
-\epsilon\bar\eta[\Gamma,\Gamma^{AB}]P_-D_A(\hat\Omega)\psi_B.
\end{eqnarray}
The last two terms are four-fermi contributions which arise from the
modification to the projection operator $\tilde P_+=P_++\epsilon\Gamma P_-$.

We make use of the Fierz identity
\begin{equation} 
\bar\eta\Gamma_B\chi\,D_A(\bar\psi_C\Gamma^{ABC}\chi)
-\bar\eta\Gamma^{ABC}\chi\,D_A(\bar\psi_B\Gamma_C\chi)
=4\bar\eta[\Gamma,\Gamma^{AB}]P_-D_A\psi_B+
2\bar\eta[(D_A\Gamma),\Gamma^{AB}]P_-\psi_B
\end{equation}
to obtain the supersymmetric variation of the action
\begin{eqnarray}
\delta S&=&{2\over\kappa^2}\int_{\cal \partial M}dv
\left(\delta g_{AB}\,\left(p^{AB}+\frac12(K^{AB}-Kg^{AB})-
\left(\hat F^{CA}\hat F_C{}^B-
\frac14g^{AB}\hat F^{CD}\hat F_{CD}\right)\right)
\right.\nonumber\\
&&+\delta C_{ABC}\left(p^{ABC}-\hat
G^{ABCN}\right)-\frac{\sqrt{2}}{96}\bar\eta\Gamma^{ABCDE}\psi_A
\left(\hat G_{BCDE}+3\sqrt{2}\hat F_{AB}\hat F_{CD}\right)\nonumber\\
&&\left.
+\delta r_{AB}q^{AB}
+\frac12\epsilon\bar\eta\{\Gamma^{AB},(D(\hat\Omega)\Gamma)\}\psi_B
+\frac12\epsilon\bar\eta\Gamma_A\chi\,D_B(\hat{\cal K})\hat F^{AB}
\right). \label{bigvars}
\end{eqnarray}
The field equations (\ref{gravfe}-\ref{chife}) and the boundary
condition (\ref{gbc}) can now be used. Only two types of four fermi terms
remain, `$\eta\psi\chi D\chi$' and `$\eta\chi\psi\psi$'. It is relatively
straight forward to show that the `$\eta\psi\chi D\chi$' terms cancel. The
`$\eta\chi\psi\psi$' terms are much more complicated and remain to be
investigated fully.

%%%%%%%%%%%%%%%%%%%%%%%%%%%%%%%%%%%%%%%%%%%%%
\bibliography{super.bib,books.bib}

%%%%%%%%%%%%%%%%%%%%%%%%%%%%%%%%%%%%%%%%%%%%%%

\end{document}